\documentstyle[prb,aps,preprint]{revtex}

\begin{document}
\draft
\tighten

\title{A ``fast growth'' method of computing free energy 
differences}

\author{D. A. Hendrix$^a$, C. Jarzynski$^b$}
\address{~ \\ (a) Physics Department,
                University of California, 
                Berkeley, CA \\
         {\tt dhendrix@socrates.berkeley.edu}}
\address{(b) Theoretical Division, Los Alamos National
                Laboratory,
             Los Alamos, NM \\
         {\tt chrisj@lanl.gov} \\
         {\rm http://qso.lanl.gov/$\sim$chrisj/}} 

\maketitle

\begin{abstract}
Let $\Delta F$ be the free energy difference
between two equilibrium states of a system.
An established method of numerically computing $\Delta F$ 
involves a single, long ``switching simulation'', during 
which the system is driven reversibly from one state to 
the other ({\it slow growth}, or {\it adiabatic switching}).
Here we study a method of obtaining the same result
from numerous independent, {\it irreversible} simulations 
of much shorter duration ({\it fast growth}).
We illustrate the fast growth method, computing the
excess chemical potential of a Lennard-Jones fluid
as a test case, and we examine the performance
of fast growth as a practical computational tool.
\end{abstract}

\pacs{\\ LAUR-00-2050}

\section*{Introduction}
\label{sec:intro}

The development of numerical methods of computing the 
free energy
difference $\Delta F=F^B-F^A$ between two equilibrium
states of a system, $A$ and $B$, is a field of active 
research\cite{frenkel,activity},
of significant practical
importance in chemistry, physics, and biology.
One widely applicable method -- often called
{\it slow growth} or {\it adiabatic switching} --
utilizes a dynamical simulation during which the system
evolves, maintained at a constant temperature, as an 
external parameter $\lambda$ is slowly changed so as 
to drive the system from the initial to the final 
equilibrium state, $A\rightarrow B$.
The evolution can occur as a sequence of discrete
Monte Carlo steps, or as a time-continuous molecular
dynamics trajectory.
The slow growth method is often presented as a
variant of {\it thermodynamic integration}\cite{kirkwood}, 
with $\Delta F$ expressed as an integral over a continuous
sequence of equilibrium states.
However, slow growth can equally well be understood
as the {\it in silico} exploitation of a basic
result of statistical mechanics\cite{reinhardt}:
\begin{equation}
\label{eq:basic}
\Delta F = W_{\infty},
\end{equation}
which states
that the external work performed on a system during
a reversible, isothermal process connecting two
equilibrium states, is equal to the free energy
difference between them.
The slow growth method amounts to simulating
such a ``switching process'' numerically,
then equating the work performed with
the desired free energy difference.

The subscript on $W_{\infty}$ emphasizes that the
process carrying the system from $A$ to $B$ is
reversible, hence (in principle) of infinite
duration.
During such a process, the system evolves through
a continuum of equilibrium states
connecting $A$ to $B$.
In practice, exact reversibility is never
achieved, and one uses a finite-time
simulation to estimate the free energy difference:
\begin{equation}
\label{eq:slowgrowth}
\Delta F \approx W_{\tau},
\end{equation}
where $W_{\tau}$ denotes the work performed during
a simulation of duration $\tau$.
The accuracy of a slow growth computation depends
on how close one is to the reversible limit;
for short switching times $\tau$, the system can 
be driven far from equilibrium, resulting in a poor
estimate of $\Delta F$.

In recent years, the following
{\it non-equilibrium work relation}
has been derived\cite{iden,mast}, suggesting an
alternative prescription for computing $\Delta F$:
\begin{equation}
\label{eq:noneq}
\exp(-\beta\Delta F) =
\overline{\exp(-\beta W_{\tau})}.
\end{equation}
This result pertains to an {\it ensemble} of
simulations of a switching process of
{\it arbitrary} duration $\tau$.
The simulations are carried out independently,
with initial conditions for the 
system sampled randomly from the canonical ensemble 
corresponding to the equilibrium state $A$.
The work $W_\tau$ is computed for each simulation,
and the overbar on the right denotes an average
over the ensemble.
The factor $\beta$ is the inverse temperature of
the initial equilibrium state, and also of the 
thermostat (see Section \ref{sec:review} below).
Eq.\ref{eq:noneq} can be rewritten as:
\begin{equation}
\label{eq:dfwx}
\Delta F =
\lim_{N\rightarrow\infty}
W_\tau^{x,N},
\end{equation}
where
\begin{equation}
\label{eq:expav}
W_\tau^{x,N} \equiv
-\beta^{-1}
\ln\Biggl[
{1\over N} \sum_{n=1}^N
\exp(-\beta W_{\tau,n})\Biggr].
\end{equation}
Here, $W_{\tau,n}$ is the work performed
during the $n$'th of $N$ independent simulations.
We will refer to $W_{\tau}^{x,N}$ as the
{\it exponential average} of $W_{\tau}$ over the
$N$ simulations (in contrast with the ordinary
average, which we will later introduce as
$W_{\tau}^{a,N}$).
We stress that the non-equilibrium work
relation is valid for arbitrary $\tau$.
This is perhaps surprising:
even for short switching
times -- for which the system is driven
far from equilibrium, and the work performed
during a typical simulation gives a sorry
estimate of the free energy difference --
we can still compute $\Delta F$
from the set of work values 
$\{W_{\tau,1},W_{\tau,2},\cdots,W_{\tau,n},\cdots\}$
obtained from an ensemble of simulations.

Since $W_{\tau}^{x,N}\rightarrow\Delta F$
as $N\rightarrow\infty$ (with $\tau$ fixed),
we are led to the following large-$N$ 
approximation:
\begin{equation}
\label{eq:fastgrowth}
\Delta F \approx W_{\tau}^{x,N}.
\end{equation}
Thus, after running many simulations of 
a given switching process,
we can estimate $\Delta F$ as the
exponential average of the values 
of work performed.
We will refer to this proposed method of
computing free energy differences as the
{\it fast growth} method, to emphasize
that it remains valid even for short
switching times $\tau$.

The aim of this paper is a
study of the fast growth method.
We have performed numerous switching
simulations of the insertion of a single
particle into a Lennard-Jones fluid,
at various switching times $\tau$.
We will use the results of these simulations
both to present an illustration of principle,
and to examine the utility of fast growth
as a practical computational method.

This paper is organized as follows.
In Section \ref{sec:review} we briefly
review the theoretical foundation of the
fast growth method.
In Section \ref{sec:model} we introduce our
test case -- the evaluation of the excess
chemical potential of a Lennard-Jones
fluid --
and we present details of the computation.
In Section \ref{sec:results} we use our
numerical results to illustrate the validity
of fast growth.
In Section \ref{sec:comparison} we use the same
results to compare the slow and fast growth
methods, given a fixed amount of computer time.
Finally, in Section \ref{sec:linresp}
we comment on the relationship between slow
growth, fast growth, and a free energy formula
based on the linear response approximation.

\section{Theoretical basis of fast growth}
\label{sec:review}

Here we introduce notation and briefly review
the theoretical basis of the fast growth method.
The emphasis will be on a presentation of the
non-equilibrium work relation (Eq.\ref{eq:noneq})
oriented specifically  toward the numerical problem 
of computing free energy differences.
For a more physical presentation,
see Ref.\cite{iden}.

Consider a system with a finite number of degrees
of freedom, $D$.
Let ${\bf z}$
denote a point in the $2D$-dimensional
phase space of the system.
Next, let $\lambda$ denote an externally controlled 
parameter (for instance, an applied field),
and let $H_\lambda({\bf z})$ be the
parameter-dependent
Hamiltonian which gives the internal energy
of the system, in terms of its
microstate ${\bf z}$.

Now consider two equilibrium states of
the system, $A$ and $B$, defined at the
same temperature, $T$, but corresponding to
different values of the external parameter,
$\lambda^A$ and $\lambda^B$.
The partition function associated with state $A$ is
given by the integral (assumed finite)
\begin{equation}
\label{eq:partfndef}
Z^A = \int d{\bf z} 
\exp [-\beta H^A({\bf z})],
\end{equation}
where $\beta = (k_BT)^{-1}$, and $H^A\equiv H_{\lambda^A}$;
and similarly for state $B$.
The free energy difference between
the two states is then
\begin{equation}
\label{eq:dfdef}
\Delta F = F^B-F^A = 
-\beta^{-1}\ln{Z^B\over Z^A}.
\end{equation}
It is this quantity that we want to compute.

Next, suppose we choose a {\it thermostating
scheme}, a numerical prescription for generating
evolution under which our system samples
the canonical ensemble associated with any
equilibrium state $(\lambda,T)$.
Intuitively, we can view the thermostating
scheme as a way of ``mocking up'' the evolution
of a system in contact with a heat reservoir.
This evolution might take the form of a 
continuous-time trajectory, or a discrete sequence 
of Monte Carlo steps.
Examples of such schemes include the Metropolis
Monte Carlo algorithm\cite{metro}, Langevin dynamics,
the Andersen thermostat\cite{andersen}, and the Nos\' e-Hoover
thermostat\cite{nose,hoover}.
Note that the thermostating scheme is itself
parametrized by $(\lambda,T)$
(e.g.\ the Metropolis 
acceptance probability takes the form
$e^{-\beta H_\lambda}$), and might be
explicitly {\it stochastic}, relying on the
generation of random numbers to produce the
evolution.

Imagine that we now run a simulation, first
sampling the initial microstate of our 
system from the canonical ensemble corresponding to 
state $A$, then letting the system evolve in time
under the thermostating scheme, 
as the value of $\lambda$ is externally changed --
or ``switched'' -- from $\lambda^A$ to $\lambda^B$,
and the temperature of the thermostat is held
fixed at $T$.
For continuous-time evolution, let $\tau$ denote 
the duration of the simulation;
let $\lambda(t)$ specify the (externally
imposed) parametric time-dependence, from
$\lambda(0)=\lambda^A$ to $\lambda(\tau)=\lambda^B$;
and consider the integral
\begin{equation}
\label{eq:wcont}
W_\tau = \int_0^\tau dt\,
\dot\lambda(t)\,
{\partial H_\lambda\over\partial\lambda}
\Bigl({\bf z}(t)\Bigr),
\end{equation}
where $\dot\lambda=d\lambda/dt$, and
${\bf z}(t)$ represents the evolution of the
system.
Alternatively, if the evolution is implemented 
with an explicitly discrete algorithm 
(e.g.\ Metropolis Monte Carlo),
then the corresponding quantity is
\begin{equation}
\label{eq:wdisc}
W_\tau = \sum_{t=0}^{\tau-1}
\Bigl[
H_{\lambda_{t+1}}({\bf z}_t) -
H_{\lambda_t}({\bf z}_t)
\Bigr],
\end{equation}
where ${\bf z}_t$ is the microstate of the
system and $\lambda_t$ the value of the
parameter, after $t$ steps;
and $\tau$ is the total number of steps in
the simulation.
In what follows, we will use the notation of
continuous-time evolution,
but the statements apply to explicitly discrete
evolution as well.

The quantity $W_\tau$ can be interpreted as
the external {\it work} performed on the system
over the course of the 
simulation.\cite{iden,reinhardt,sekimoto,zakopane,crooks,schrodinger}
We will use this interpretation, though we
warn the reader that Eq.\ref{eq:wcont} 
(or Eq.\ref{eq:wdisc}) is by no means an accepted, 
textbook definition of work, and the interpretation 
of $W_\tau$ as work may be subject to dispute.
The results presented, however, do not depend
on the validity of this interpretation.

The trajectory ${\bf z}(t)$ which emerges
from our simulation (and which in turn determines
the value of $W_\tau$)
depends on a string of random numbers:
those involved in sampling the initial conditions,
as well as those, if any, needed to generate
the evolution itself.
By repeating the simulation with different
sets of random numbers, we obtain different
{\it realizations} of the process we are
simulating.
We view these as representing independent samples
from an infinite ensemble of all possible
realizations.\cite{indepsamples}
Eq.\ref{eq:noneq} then states that
the average value of
$e^{-\beta W_\tau}$ over the ensemble
of possible realizations is
equal to $e^{-\beta\Delta F}$,
with $\Delta F$ as given by Eq.\ref{eq:dfdef}.
In other words, 
$e^{-\beta W_\tau}$ is an {\it unbiased
estimator} of $e^{-\beta\Delta F}$.
By collecting sufficiently many samples from
the ensemble of realizations (i.e.\ by repeatedly 
running simulations) we can in principle compute
the average of $e^{-\beta W_\tau}$ 
to the desired degree of accuracy. 
This is the essence of the fast growth method.

For a variety of thermostating schemes, 
Eq.\ref{eq:noneq} has been established as an
exact result, 
valid for any finite switching time $\tau$.
Precise conditions under which this result
is rigorously true are presented in Refs.\cite{mast,crooks}.
In particular, all four thermostating schemes
mentioned earlier -- Metropolis, Langevin,
Andersen, Nos\' e-Hoover -- satisfy these
conditions.

Note that we do not automatically gain something
for nothing with the fast growth method:
while it allows us to compute
$\Delta F$ without the bother of running a
very long simulation, the price paid is
that many short simulations might be needed.

Finally, we mention that, in the limit $\tau\rightarrow 0$,
Eq.\ref{eq:noneq} reduces to the well-known
{\it free energy perturbation} identity\cite{zwanzig}:
\begin{equation}
\label{eq:fep}
\exp(-\beta\Delta F) =
\langle\exp(-\beta\Delta H)\rangle_A,
\end{equation}
where $\Delta H\equiv H^B-H^A$ is the difference
between the two Hamiltonians, evaluated at a given
microstate;
and $\langle\cdots\rangle_A$ denotes an average over
microstates sampled from the canonical ensemble $A$.

\section{Test case: excess chemical potential of 
a Lennard-Jones fluid}
\label{sec:model}

To illustrate the fast growth method,
we chose to compute the excess chemical potential
of a three-dimensional Lennard-Jones fluid.
This is equal to the free energy change associated
with ``turning on'' the interactions between a
single particle and the remainder of the
fluid, at constant volume and temperature\cite{frenkel}.
This process can be viewed as the transfer of one
particle from an ideal gas to the Lennard-Jones
fluid; the {\it excess} chemical potential is 
defined as the chemical potential of the fluid 
relative to that of the ideal gas.
Following the notation of Mon and Griffiths\cite{mg},
let $U_M$ denote the potential energy of $M$
identical, classical particles interacting
pairwise:
\begin{equation}
U_M({\bf r}_1,\cdots,{\bf r}_M) =
\sum_{i<j} \Phi(\vert {\bf r}_j-{\bf r}_i\vert),
\end{equation}
where $\Phi$ is the interaction potential.
Now imagine a system of $M+1$ particles, and
consider two Hamiltonians:
\begin{eqnarray}
H^A({\bf z}) &=&
\sum_{i=1}^{M+1} {p_i^2\over 2m} + 
U_M({\bf r}_1,\cdots,{\bf r}_M) \\
H^B({\bf z}) &=&
\sum_{i=1}^{M+1} {p_i^2\over 2m} + 
U_{M+1}({\bf r}_1,\cdots,{\bf r}_{M+1}).
\end{eqnarray}
In both cases ${\bf z}$ is the
$6(M+1)$-dimensional vector specifying the
positions and momenta of all the particles.
$H^A$ describes $M$ interacting
particles, plus a ``tagged'' particle
moving freely;
$H^B$ describes $M+1$ interacting
particles.
We take the positions ${\bf r}_i$
to be confined within a cube of volume $V=L^3$
with periodic boundary conditions.
Note that the difference
\begin{equation}
\Delta H \equiv H^B-H^A =
\sum_{i=1}^M \Phi(\vert {\bf r}_{M+1}-{\bf r}_i\vert)
\equiv \Psi({\bf r}_1,\cdots,{\bf r}_{M+1})
\end{equation}
is just the interaction energy between the tagged
particle and the others.

At a given temperature $T$, Eqs.\ref{eq:partfndef}
and \ref{eq:dfdef} define partitions functions
$Z^A$ and $Z^B$, and the free energy difference
$\Delta F$, for the Hamiltonians $H^A$ and $H^B$.
This free energy difference, i.e.\ the work
associated with reversibly turning
on the interaction $\Psi$ at constant $V$ and $T$,
is the excess chemical potential:
\begin{equation}
\mu_{ex}=\Delta F
\end{equation}

Let us now parametrize a {\it family} of
Hamiltonians, $H_\lambda({\bf z})$, interpolating
linearly from $H^A$ to $H^B$:
\begin{equation}
H_\lambda = H^A +
\lambda\Psi,
\end{equation}
so that $H^A=H_0$ and $H^B=H_1$.
The intermediate Hamiltonians ($0<\lambda<1$)
represent a partially interacting tagged
particle.

A switching simulation then proceeds as follows.
We start by sampling initial conditions from the
canonical ensemble corresponding to $H^A$.
We then allow the $(M+1)$-particle system to
evolve in time under a chosen thermostating
scheme, as $\lambda$ is switched from 0 to 1
over a time $\tau$, i.e.\ from a non-interacting
to a fully interacting tagged particle.
By running either a single simulation with
$\tau$ very large (slow growth), or many
simulations with smaller $\tau$ (fast growth),
we can construct an estimate of $\mu_{ex}$ 
from the value(s) of external work, $W_\tau$,
performed during the simulation(s), using
Eq.\ref{eq:slowgrowth} or \ref{eq:fastgrowth}.

At the end of Section \ref{sec:review},
we alluded to the fact that,
in the limit $\tau\rightarrow 0$,
fast growth reduces to the free energy perturbation
method of computing $\Delta F$.
For the particular case considered here, this is
equivalent to Widom's {\it particle insertion method}\cite{frenkel,widom},
which gives $\mu_{ex}$ in terms of sampled values of
the energy cost ($\Delta H$) of suddenly materializing
an additional particle at a randomly chosen location
within the fluid.

We performed switching simulations
at various switching times $\tau$.
The results of these will be presented in
the following section.
Here we describe the details and parameters
of the simulations themselves.

The usual Lennard-Jones interaction potential,
for a pair of particles separated by a distance
$r$, is given by
\begin{equation}
\Phi_{LJ}(r) = 4 \epsilon 
\left[ \left( {\sigma\over r} \right)^{12}-
\left( {\sigma\over r} \right)^6 \right],
\end{equation}
where $\epsilon$ is the depth of the potential at its minimum,
and $\sigma$ is the van der Waals diameter of the particle.
We instead used a modified potential:
\begin{equation}
\Phi(r) = \left\{ \begin{array}{ll}
a - b r^2 
& \qquad 0\le r \le 0.8\sigma\\
\Phi_{LJ}(r)+
c(r - r_c)-d
& \qquad 0.8\sigma \le r \le r_c \\
0
& \qquad r_c \le r.
\end{array}
\right.
\end{equation}
Here $r_c$ is a cutoff distance, which we took
to be half the length of a side of the box:
\begin{equation}
r_c = L/2.
\end{equation}
and the constants $a$, $b$, $c$, and $d$ were
chosen to preserve the continuity of $\Phi$
and its derivative.
We used $\Phi(r)$ rather than $\Phi_{LJ}(r)$
both to avoid the singularity at $r=0$, and
to impose a strict radius of interaction
equal to $L/2$.
The need for the latter condition arose
from our use of periodic boundary conditions,
which is equivalent to an infinite lattice of 
identical cells:
the requirement that $\Phi$ vanish for $r>L/2$
guarantees that, given a pair of particles,
each interacts only with a single image of
the other.

For the parameter values we have chosen, the
difference between $\Phi(r)$ and $\Phi_{LJ}(r)$
is very small for $r>0.8\sigma$.
Furthermore, we have found the core repulsion between
particles to be sufficiently strong that two fully
interacting particles will essentially never
come closer than a distance $0.8\sigma$ of one
another.
These considerations suggest that the excess chemical
potential $\mu_{ex}$ which we plan to compute
will differ little from that defined for the usual
Lennard-Jones fluid.
However, this issue is not particularly
relevant for the central aim of this paper.

Since the Lennard-Jones potential works especially 
well with inert gases, we modeled an argon fluid.  
The parameters associated with argon are\cite{reid} 
\begin{eqnarray}
\sigma &= 3.542 \times 10^{-10} {\rm m} &~=~ 3.542~{\rm \AA} \\
\epsilon &= 1.288 \times10^{-21} {\rm J} &~=~ 0.1854~{\rm kcal/mol} \\
m &= 6.634 \times 10^{-26} {\rm kg} &~=~ 39.94~{\rm amu}.
\end{eqnarray}
We took these values to define the units of
length, energy, and mass in our simulation.
The units for time and temperature are then:
$\sigma\sqrt{m/\epsilon} = 2.542 {\rm ps}$
and
$\epsilon/k_B = 93.3{\rm K}$, respectively.
Henceforth, all values will be quoted in these
units.

We chose the 
{\it Andersen thermostat}\cite{andersen}
to simulate the effects of an external heat reservoir.
In this scheme, the evolution of the fluid
is governed by Newton's equations, supplemented
by the forced thermalization of the momenta
of randomly chosen particles, at regular intervals.
Specifically, at fixed time intervals
$\delta t_{\rm And}$, a particle is chosen
at random, and all three components
of its momentum vector are replaced by
{\it thermal} values,
sampled randomly from the Maxwell-Boltzmann
momentum distribution at the thermostat 
temperature $T$.
To carry out the integration of Newton's
equations, we used the velocity-Verlet 
algorithm\cite{frenkel}.

All of our simulations were run using $M=125$
untagged particles, plus one tagged particle,
inside a box of sides $L=5.3$, with the thermostat
set at $T=1$.

During each switching simulation
(following a relaxation interval at $\lambda=0$,
to generate an initial equilibrium microstate;
see Section \ref{sec:comparison}), the interaction
strength of the untagged particle was turned on
according to the schedule:
\begin{equation}
\lambda(t) = (t/\tau)^2,
\end{equation}
from $t=0$ to $t=\tau$.
The time step for the velocity-Verlet integration 
was taken to be $\delta t_{\rm Ver}=.01$, and
the Andersen momentum thermalization was
implemented at every time step:
$\delta t_{\rm And}=.01$.

\section{Results: illustration of principle}
\label{sec:results}

We ran a varying number of simulations of 
the switching process, at seven 
different values of the switching time, from
$N=3334$ simulations at $\tau=3.0$, to
$N=10$ simulations at $\tau=1000.0$,
keeping the total switching time devoted to
each set of simulations, $N\tau$, 
fixed at (approximately) $10^4$.
For every simulation, we computed the work
performed as a result of turning on the
interactions between the tagged particle
and the rest.
We then calculated 
both the ordinary and exponential average of these
values of work, over the simulations 
corresponding to each value of
$\tau$:
\begin{eqnarray}
W_\tau^{a,N} &=&
{1\over N}\sum_{n=1}^N W_{\tau,n}\\
W_\tau^{x,N} &=&
-\beta^{-1}\ln
\left[
{1\over N}\sum_{n=1}^N
\exp(-\beta W_{\tau,n})
\right],
\end{eqnarray}
where $W_{\tau,n}$ denotes the work performed
during the $n$'th simulation of duration $\tau$.

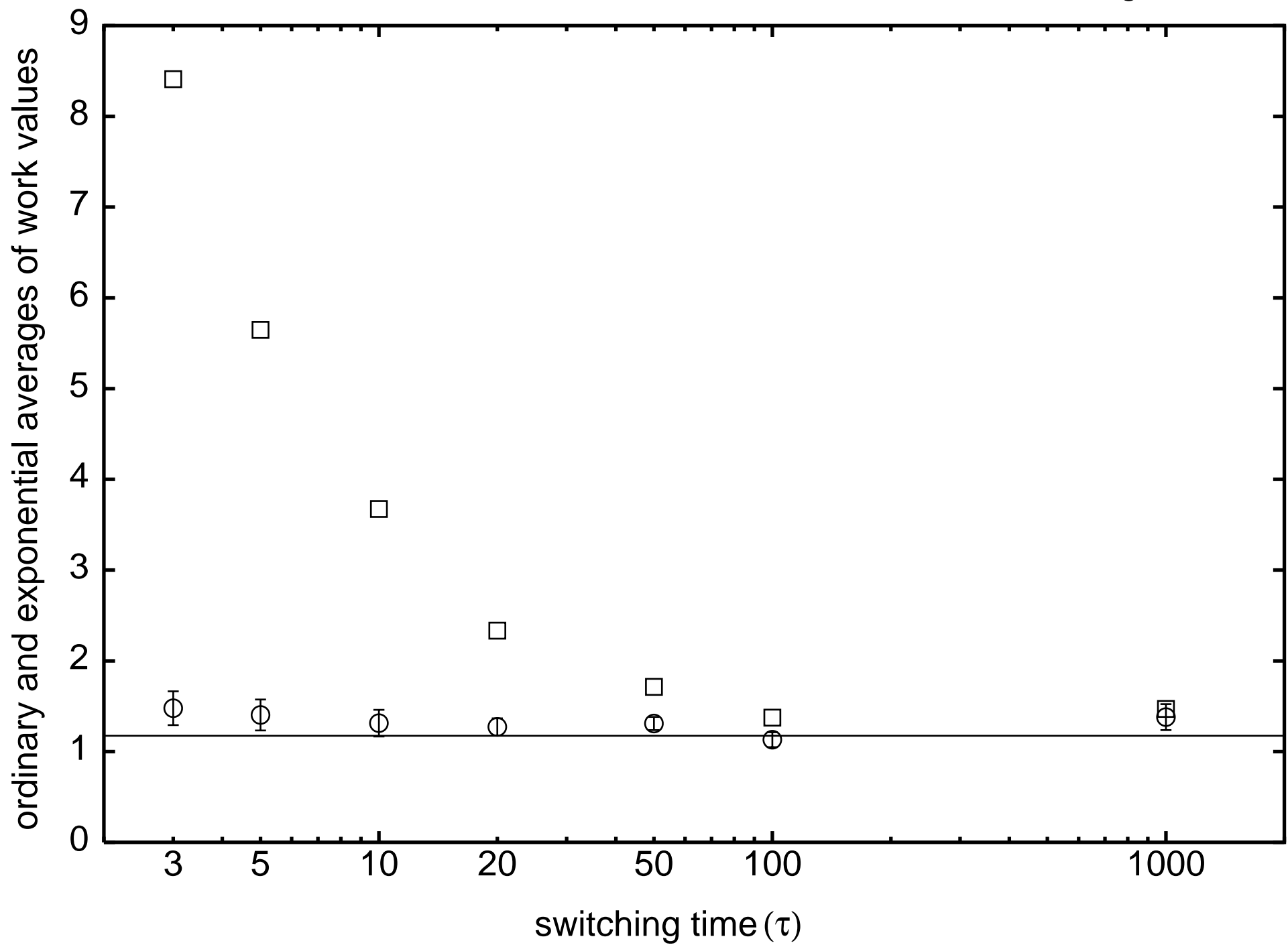
\begin{figure}
\caption{
For each of the seven sets of simulations,
corresponding to switching times from
$\tau=3.0$ to 1000.0, we computed both
the ordinary average ($W^a$) and the
exponential average ($W^x$) of the work
values.
These are shown in the figure, along with
our estimate for the true value of
$\Delta F$, the solid line at $W=1.174$.
The ordinary averages, $W^a$, displayed as
squares,
over-estimate $\Delta F$, although the 
difference decreases as one approaches the 
reversible limit.
By contrast, the
exponential averages $W^x$ (circles) provide 
significantly better estimates of $\Delta F$.
The number of simulations, $N$, performed for
each value of the switching time, $\tau$,
was chosen so that $N\tau=10000.0$ in each
case (except $N\tau=10002.0$ for $\tau=3.0$).}
\label{fig:principle}
\end{figure}

The results are summarized in Fig.\ref{fig:principle}.
The error bars on the exponential averages (circles)
were computed with the bootstrap method\cite{bootstrap}.
Error bars on the ordinary averages (squares) are not
shown, but were typically the size of the squares
themselves.

Additionally, we performed two very long
simulations at $\tau=25000.0$, and used the
average of these as our best estimate of the
true value of $\Delta F$
(defined by Eq.\ref{eq:dfdef}).
This is shown as the horizontal line at
$W=1.174$ in Fig.\ref{fig:principle}, and
-- judging from the difference between these
two (nearly quasi-static) simulations --
this value should be understood to have an
uncertainty of $\sim 0.1$.

Let us now consider the salient features of
Fig.\ref{fig:principle}, beginning 
with the values of $W_{\tau}^{a,N}$.
If we view the work performed during a single
simulation, $W_{\tau}$, to be an estimate of
the free energy difference $\Delta F$,
then this estimate is subject to both
statistical and systematic errors\cite{mm,wood,hermans}.
The former reflect the {\it spread} in the
distribution of $W_{\tau}$ values (for a given
$\tau$), the latter a {\it bias}:
$W_{\tau}$ tends to over-estimate $\Delta F$.
This bias follows rigorously from the
non-equilibrium work relation\cite{iden},
but can also be understood in terms of 
thermodynamic considerations\cite{reinhardt,tsao}
(e.g.\ the Second Law), if one accepts the
interpretation of $W_\tau$ as work.
By averaging $W_\tau$ over infinitely many 
realizations, the statistical error is removed, 
but the systematic error remains:
\begin{equation}
\lim_{N\rightarrow\infty} W_{\tau}^{a,N}
= \overline{W_{\tau}}>\Delta F
\qquad ({\rm finite}\,\,\tau).
\end{equation}
Thus, for any value of $\tau$, the quantity
$\overline{W_{\tau}}-\Delta F$ 
represents the {\it systematic error} associated
with using simulations of duration $\tau$
to estimate $\Delta F$.
As expected, this error (approximated by
$W_{\tau}^{a,N}-\Delta F$ in Fig.\ref{fig:principle})
is positive, but decreases as we approach the 
reversible limit, $\tau\rightarrow\infty$.

The {\it exponential average} of $W_{\tau}$ over $N$
simulations of duration $\tau$, viewed as
an estimate of $\Delta F$, also contains
errors when $N$ is finite\cite{mast},
but these are removed
in the limit of infinitely many simulations:
\begin{equation}
\lim_{N\rightarrow\infty} W_{\tau}^{x,N}
=\Delta F
\qquad ({\rm arbitrary}\,\,\tau).
\end{equation}
The values of $W_{\tau}^{x,N}$ shown in
Fig.\ref{fig:principle} are consistent with this 
statement.
All of the exponential averages closely approximate 
(our best estimate of) the free energy difference,
$\Delta F=1.174$, illustrating the validity of the 
fast growth method.

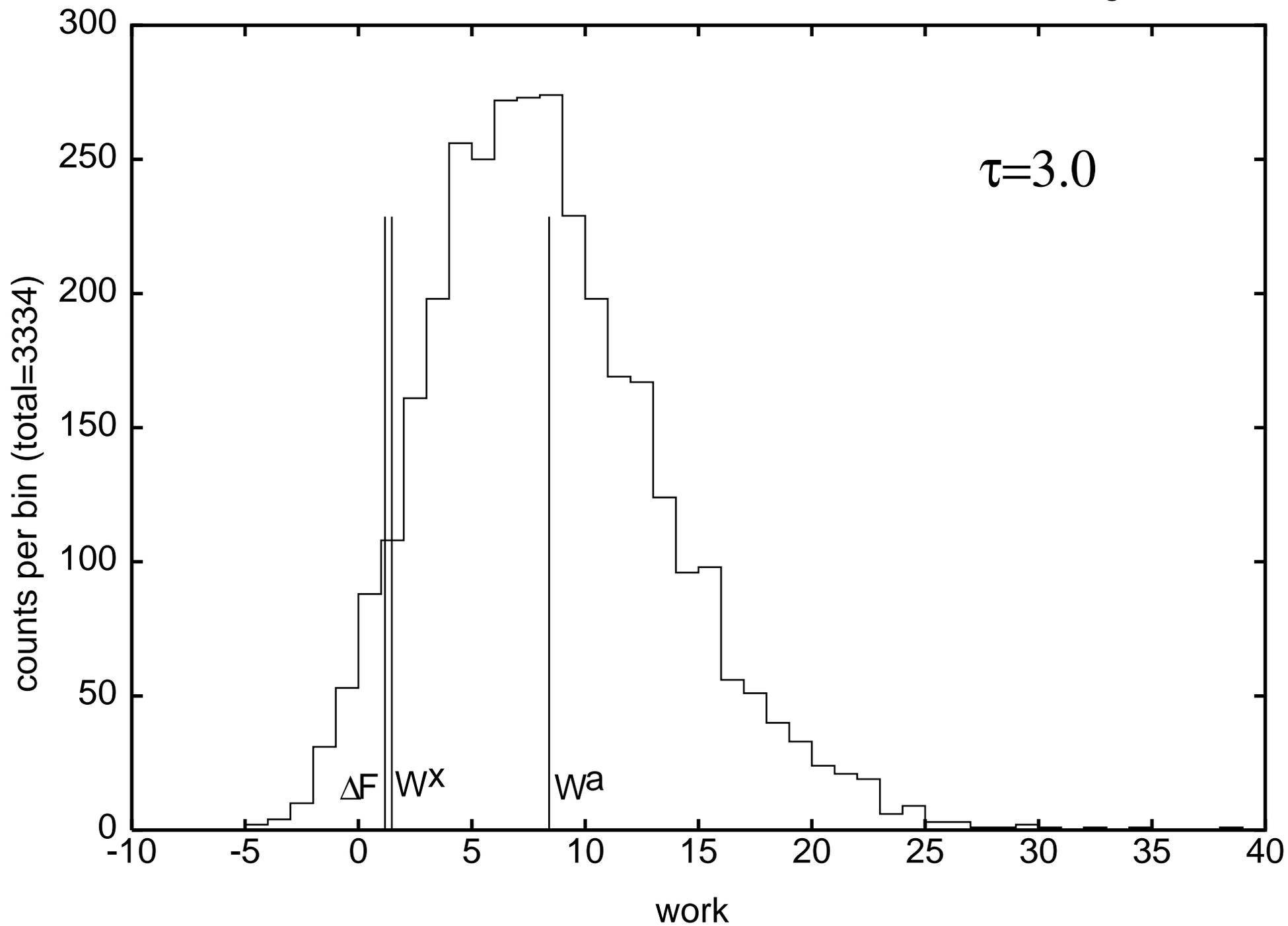
\begin{figure}
\caption{
A histogram, showing the distribution
of values of $W$, in bins of
unit energy width, for
the 3334 simulations performed at the
shortest switching time, $\tau=3.0$.
The three vertical lines show the
ordinary average of these work values,
$W^a$,
the exponential average, $W^x$, and 
(our best estimate of) the free energy
difference, $\Delta F$.}
\label{fig:hist}
\end{figure}

The histogram in Fig.\ref{fig:hist} displays
the distribution of work values for the
set of simulations performed at $\tau=3.0$.
The three vertical lines show
the locations of the ordinary average of the
work values, $W^a=8.413$,
the exponential average, $W^x=1.478$,
and the free energy difference, 
$\Delta F = 1.174$,
shown in Fig.\ref{fig:principle}.
(Throughout the paper, we frequently suppress
the subscript $\tau$ and/or the superscript $N$.)
Note that $W^x$ and $\Delta F$
are quite close: despite values of $W$
ranging from $-4.3$ to $+34.1$, the exponential
average $W^x$ differs from $\Delta F$
by only $\sim 0.3$, which is not much larger than
our uncertainty in the value of $\Delta F$ itself!

Figs.\ref{fig:principle} and \ref{fig:hist}
are intended as an ``illustration of principle'' 
of the fast growth method,
demonstrating that the value of $\Delta F$
can be extracted from an ensemble of finite-time
switching simulations, during which the system
is explicitly driven away from equilibrium.

\section{Comparison of methods}
\label{sec:comparison}

Both slow
and fast growth converge to the correct value
of $\Delta F$ in the limit of infinite
computational resource:
$\tau\rightarrow\infty$ for slow growth,
$N\rightarrow\infty$ for fast growth.
In practice, of course, we can perform neither
a single infinitely long simulation, nor infinitely
many finite-time simulations, so it becomes
interesting to pose the following question:
Given a {\it finite} amount of computer time,
which method is preferable?
We now consider this question empirically, 
in a manner relevant to the practical computation
of free energy differences.
Specifically, we study 
how the accuracy of the predicted free energy
difference depends on the choice of the switching
time, $\tau$, given a fixed amount of total
computational time.

Imagine that we have been allotted
$\tau_{\rm TOT}$ units of simulation time.
In this total we must include not only
the time devoted to repetitions of the switching
process itself, but also the time used to generate
a microstate sampled from equilibrium, before the
start of each process.
If each switching realization is preceded by
a relaxation interval of duration $\tau_{\rm REL}$
(to generate the initial microstate), then the
number of switching realizations possible is given
by the largest integer $N$ satisfying the
inequality:
\begin{equation}
\label{eq:N}
N \le 
{\tau_{\rm TOT}\over \tau_{\rm REL}+\tau}.
\end{equation}

In the simulations described in Section \ref{sec:results},
for each value of $\tau$ the results were obtained
from a single, long run, in which realizations
of the switching process itself alternated with
relaxation intervals of duration
$\tau_{\rm REL}=1.0$, during which the
system evolved at fixed $\lambda=0$.
(The final microstate at the end of a given
switching simulation was taken as the seed
for the following relaxation 
interval.\cite{initialmicrostate}
The value of $\tau_{\rm REL}$ was chosen on
the basis of simulations
showing that correlations in our system decay
to zero over $\sim 0.6$ time units.)
We used these results to investigate predictions
of $\Delta F$ for various switching times $\tau$,
given a total resource of $\tau_{\rm TOT}=1050.0$
time units of simulation.
For each of the seven switching times
considered, we generated ten predictions of
$\Delta F$ using the data from the simulations.
For instance, at $\tau=5.0$, Eq.\ref{eq:N} gives
$N=175$; therefore, of the 2000 switching
realizations performed at $\tau=5.0$, we used the 
first 1750 to generate ten independent estimates of 
$\Delta F$.
The remaining 250 realizations were not used here.

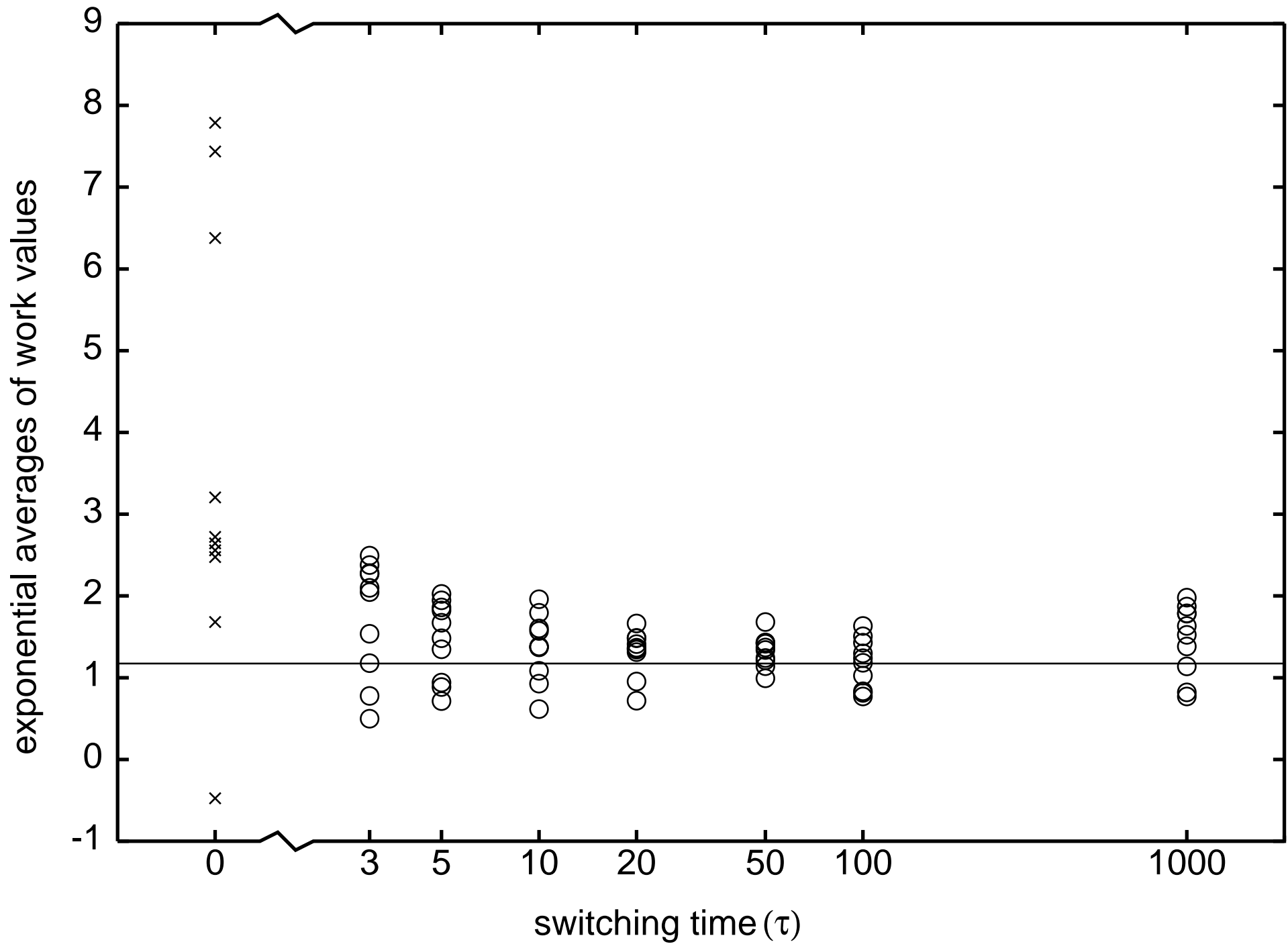
\begin{figure}
\caption{
Each circle or cross represents an estimate of 
$\Delta F$, using no more than 1050.0 units of
simulation time.
The ten circles at $\tau=1000.0$ are thus
slow growth estimates;
all other circles are fast growth estimates;
and the crosses at $\tau=0.0$ were obtained 
with the Widom method.}
\label{fig:comparison}
\end{figure}

Fig.\ref{fig:comparison} displays the results, 
for the various switching times.
Each circle represents an estimate of $\Delta F$
obtained from a total of (no more than) $1050.0$ units
of simulation time.
The data at $\tau=1000.0$ are
slow growth estimates ($N=1$), whereas the
remaining points make explicit use of the
fast growth formula (Eq.\ref{eq:expav}).
A striking feature of these results is that the
accuracy of the predictions does not seem to depend
drastically on $\tau$.
In other words, for the example
considered in this paper, {\it fast growth and slow growth
yield comparably good estimates of $\Delta F$
for a fixed computational resource, over a wide
range of switching times}.

Additionally, we generated ten estimates of
$\Delta F$ using the Widom method --
shown as crosses at $\tau=0.0$ --
again with a total of $1050.0$ time units devoted 
to each estimate.
Here, however, we effectively took $\tau_{\rm REL}=0.1$:
after every time interval $\Delta t=0.1$,
we generated a random location for the $(M+1)$'th
particle, determined $\Delta H$, and then continued
with the simulation of $M=125$ mutually interacting
particles.
Thus, $10500$ samples of $\Delta H$ contributed to
each of the ten Widom estimates.
As we made no attempt to optimize $\Delta t$,
these results do not really represent a direct
and fair comparison between fast growth and the
Widom method, and rather serve as an illustration
of the latter.

In the asymptotic limit of large $N$ and
large $\tau$, Crooks\cite{crooksthesis} has shown
that the size of errors in the estimate of
$\Delta F$ depends only on $N\tau$.
On the other hand, for fast switching, 
the distribution of
work values can become quite wide, resulting in slow 
convergence of $W_{\tau}^{x,N}$ with $N$.
This is a well-known problem associated with
computing exponential averages (as with the free energy
perturbation method\cite{wood}),
which manifests itself both in large statistical errors,
and -- if the distribution is wide enough -- as
a significant bias toward over-estimation of $\Delta F$.
The Widom data ($\tau=0.0$) certainly shows these
symptoms; the somewhat wide scatter of predictions
at $\tau=3.0$ is likely also evidence of this problem.

We suspect that, generically, the expected error 
associated with fast growth decreases with increasing
$\tau$ (for fixed $\tau_{\rm TOT}$).
If this is indeed the case, then Fig.\ref{fig:comparison}
suggests that for $\tau\ge 5.0$ this
trend is lost in the noise;
any gain in accuracy obtained by using
one switching realization of duration $1000.0$ rather 
than 175 realizations of duration $5.0$, is likely
to be smaller than the unavoidable statistical error 
associated with trying to compute $\Delta F$ with a 
finite resource of $1050.0$ units of simulation time.

The above, tentative conclusion -- that for a given
$\tau_{\rm TOT}$ the accuracy does 
not depend strongly on $\tau$, over a wide range of
switching times --
suggests a certain advantage of fast growth over slow
growth as a practical computational tool.
Namely, fast growth allows for the immediate estimation
of errors\cite{crooksthesis},
both statistical and systematic.
Using the $N$ values of work obtained from
different switching realizations, the
statistical uncertainty in the average of 
$e^{-\beta W_\tau}$ -- the unbiased estimate
of $e^{-\beta\Delta F}$ -- can be ascertained in the
usual way, in terms of the variance;
the resulting upper and lower bounds defining
the error bar then translate immediately into
lower and upper bounds on the estimate of
$\Delta F$ itself.
Alternatively, one can use the bootstrap
method\cite{bootstrap}, which is sensitive to the possibility
that the exponential average might be dominated
by one or a few particularly small values of work, 
out of the $N$ values obtained from the 
switching realizations.
Slow growth, by contrast, produces
only a single value of work, which by itself
provides no estimate of its own error!

In addition to statistical errors, the fast growth
estimate contains a systematic bias:
for finite $N$, $W_\tau^{x,N}$ tends to
over-estimate $\Delta F$, though to a lesser degree
than $W_\tau^{a,N}$\cite{mast}.
This bias has the same origin as the
``sample-size hysteresis'' studied in the context
of the free energy perturbation method,
and Eq.10 of Ref.\cite{wood} gives us a leading-order
estimate of its size:
\begin{equation}
\label{eq:expavsys}
\overline{W_\tau^{x,N}} - \Delta F
\approx \beta \sigma_W^2/2N.
\end{equation}
Here the overbar denotes an expectation value
(i.e.\ the average over infinitely many attempts to
estimate $\Delta F$ with $N$ values of $W$, using
fast growth), and $\sigma_W^2$ is the variance in
the work values.
Note that this bias vanishes as $N\rightarrow\infty$,
as demanded by consistency with Eq.\ref{eq:dfwx}.

For each of the fast growth estimates shown in 
Fig.\ref{fig:comparison}, we have estimated both the
statistical uncertainty (using the bootstrap method)
and the systematic bias (using Eq.\ref{eq:expavsys}).
We found that the latter never exceeded 0.07, and that
the estimated statistical uncertainty was typically about 
an order of magnitude greater than the estimated bias.
These conclusions are in agreement with
Fig.\ref{fig:comparison}, which visually suggests
that statistical rather than systematic errors dominate
our fast growth predictions.

With fast growth, the easy estimation of error bars
can furthermore be harnessed to determine,
``on the fly'', the optimal amount of computer time
to devote to the estimation of $\Delta F$:
the user simply performs one simulation
after another, updating both $W_{\tau}^{x,N}$
and its error bar after each simulation, and stops
when the size of the error becomes satisfactorily
small.
With slow growth, one must estimate ahead of time
how long the simulation needs to be, given the
desired accuracy.

Our discussion so far has been implicitly oriented
toward the computation of free energy differences
on {\it single-processor} computers, where easy
estimation of statistical errors seems to be the
main advantage of fast growth over slow growth.
However, on parallel machines another considerable
advantage emerges: fast growth is
{\it embarrassingly parallelizable}.
Given $P$ processors and a finite amount of
dedicated run time, the user simply lets each
processor perform one or more switching
simulations, and at the end all the values of
work are gathered and the exponential average
is computed.
This provides the maximal efficiency
of parallelization, $\sim 100\%$, as there is
essentially no ``cross-talk'' between processors. 
Furthermore, the additional programming effort
is minimal: once a code has been written to
perform switching simulations on a single 
processor, little more needs to be done to let 
$P$ copies of the code run independently on $P$
processors.
By contrast, to efficiently
distribute a single simulation over $P$
processors requires considerable programming
skills and time, and the optimal parallelization
scheme may well depend on details of the hardware
-- e.g.\ are the processors divided into shared-memory
clusters?, how fast is the communication between
different processors?, etc. -- making it less
portable.

The fast growth method is particularly well-suited
for Beowulf-type clusters, built
at relatively low cost with off-the-shelf
hardware components.
In such clusters, the processors typically do not
share memory, and communication between them
is often relatively slow.
Both drawbacks are avoided by
any embarrassingly parallelizable method such as
fast growth.

\section{Fast growth, slow growth, and linear response}
\label{sec:linresp}

The idea of computing $\Delta F$ from numerous
values of $W_\tau$ is not new.
Indeed, it has been recognized for some time
that, in the interest of having an estimate of
statistical errors, several or more independent
switching simulations should be carried out\cite{mm}.

Given a number of independent values of $W_\tau$,
the simplest imaginable approximation of $\Delta F$
-- though clearly not the best (Fig.\ref{fig:principle}) --
is just the ordinary average of these values:
\begin{equation}
\label{eq:simplestapprox}
\Delta F \approx W_\tau^a
\end{equation}
This amounts to slow growth, with statistical errors
reduced by averaging over samples.
In 1991, Hermans\cite{hermans}
suggested instead the following formula:
\begin{equation}
\label{eq:linresp}
\Delta F \approx
W_\tau^a - \beta\sigma_W^2/2,
\end{equation}
where $\sigma_W^2$ is the variance of the set of work
values
(closely related results are found in
Wood {\it et al}\cite{wood}, and -- for isolated
systems -- in Tsao {\it et al}\cite{tsao}).
This result can be viewed as a fluctuation-dissipation
relation, and therefore ought to be valid in the
{\it linear response} regime, i.e.\ when $\lambda$
is switched slowly enough that the system remains
near equilibrium for the duration of the process.

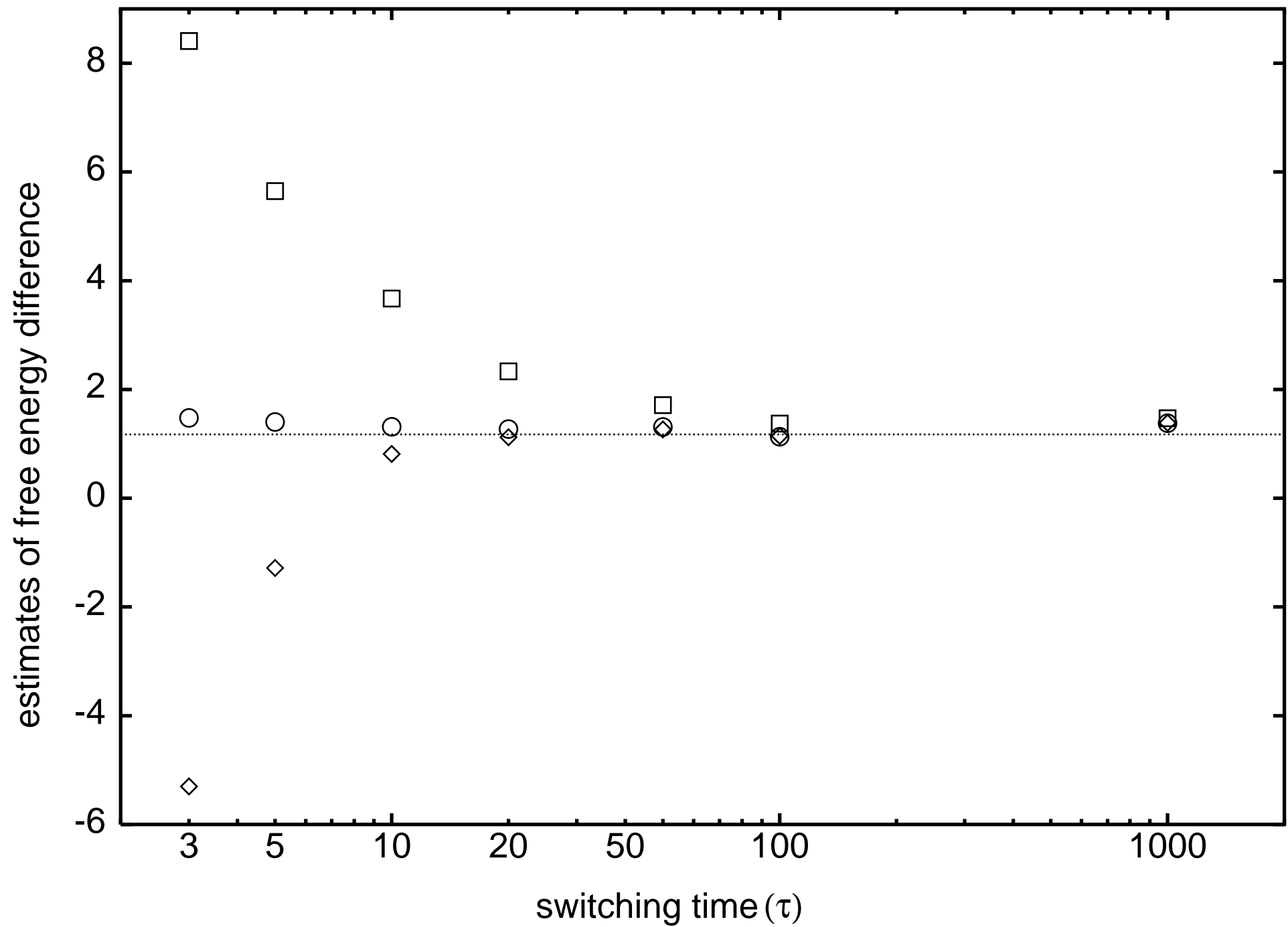
\begin{figure}
\caption{
The squares, circles, and diamonds show
three sets of estimates of $\Delta F$:
$W^a$, $W^x$, and $W^a-\beta\sigma_W^2/2$,
respectively.
These estimates were obtained from the same
data used in the previous figures, i.e.
the squares and circles are the those
shown in Fig.\ref{fig:principle}.}
\label{fig:linresp}
\end{figure}

Eq.\ref{eq:linresp} represents a
correction to Eq.\ref{eq:simplestapprox}, aiming
to compensate for the systematic
error (see Section \ref{sec:results}) 
inherent in using $W_\tau$ as an estimate of $\Delta F$.
Thus we can expect
$W_\tau^a - \beta\sigma_W^2/2$
to give a better estimate of $\Delta F$ than
$W_\tau^a$, {\it in the regime of validity of
linear response}.
Using the work values obtained from the simulations 
described in Section \ref{sec:results}, 
we have compared -- for various switching times --
three estimates of $\Delta F$:
$W_\tau^a$ (corresponding to slow growth, with
the statistical error reduced by averaging),
$W_\tau^a - \beta\sigma_W^2/2$ (linear response),
and $W_\tau^x$ (fast growth).
Fig.\ref{fig:linresp} displays the results.
As expected, there exists a near-equilibrium
regime (roughly, $\tau\ge 20.0$) in which
linear response does a good job of removing
the systematic error inherent in using $W_\tau$
as an estimate of $\Delta F$.
For shorter times, however, linear response breaks down,
as the system is driven significantly away from
equilibrium.
Fig.\ref{fig:linresp} furthermore illustrates
that the non-equilibrium work relation
is truly a {\it far-from-equilibrium} result,
remaining valid even outside the regime of validity
of the near-equilibrium prediction (Eq.\ref{eq:linresp}).

The relationship between the three estimates
considered in Fig.\ref{fig:linresp} can be
understood by rewriting Eq.\ref{eq:noneq} as
follows\cite{iden}: 
\begin{equation}
\label{eq:cumulants}
\Delta F = \sum_{n=1}^\infty
(-\beta)^{n-1}{\omega_n\over n!},
\end{equation}
where $\omega_n$ is the $n$'th cumulant
of the distribution of values of $W$.
Keeping only the first term on the right
side gives us $\Delta F\approx W_\tau^a$,
whereas keeping two terms yields
the linear response approximation,
$\Delta F\approx W_\tau^a - \beta\sigma_W^2/2$.
Therefore these approximations can be viewed
as the first two in an expansion derived from
the non-equilibrium work relation, 
Eq.\ref{eq:noneq}.

Note that one can justify keeping only the
first two terms in the cumulant expansion,
by assuming a Gaussian distribution of
work values:
since all higher ($n>2$) cumulants vanish for
a Gaussian, Eq.\ref{eq:cumulants} reduces
to Eq.\ref{eq:linresp} in that case.
More directly (i.e.\ without resorting to a
cumulant expansion), it is easily verified that
for an assumed Gaussian distribution of work values,
Eq.\ref{eq:noneq} implies a relationship between
the mean and the variance of the distribution,
namely $\Delta F = \overline W - \beta\sigma_W^2/2$.

It is easy to argue heuristically that a Gaussian
distribution of work values emerges in the linear
response regime, i.e.\ when the switching is slow enough
($\tau$ large enough) to keep the system near
equilibrium.
Namely, we can imagine that the work $W$, for one
simulation, becomes a sum of many independent
contributions:
if we break the long time interval $0<t<\tau$
into many shorter segments -- each, however,
longer than the correlation time associated with
the dynamics of the system -- then the contributions
to $W_\tau$ from the various segments will be
(roughly) independent.
The central limit theorem (CLT) then leads us
to expect that, given many realizations of this
process, the distribution of work values, $\rho(W)$, 
will be approximately Gaussian.
However, this argument requires some care:
if the spread in the values of $W$ is large
($\sigma_W\gg k_BT$), then the dominant contribution
to $\int\rho e^{-\beta W}$ comes from values
in the (lower) {\it tail} of $\rho(W)$, precisely
where the CLT ``breaks down''\cite{searles}.
So a more careful justification invoking the
CLT (to derive Eq.\ref{eq:linresp} from
Eq.\ref{eq:noneq}) requires also that
$\sigma_W$ be not much larger than $k_BT$.
This will be satisfied for sufficiently slow
switching, since
$\rho\rightarrow\delta(W-\Delta F)$
as $\tau\rightarrow\infty$ (Eq.\ref{eq:basic}).
Hence, the argument is ultimately justified.

\section*{Conclusions}

In this paper, we have studied the fast growth method, 
applying it to compute the excess chemical potential
of a (modified) Lennard-Jones fluid.
Figs.\ref{fig:principle} and \ref{fig:hist} illustrate
the validity of the method, showing that it accurately 
estimates $\Delta F$ over a wide range of
switching times, including short times for which the
system is driven far from equilibrium.
Moreover, Fig.\ref{fig:comparison} suggests that
the accuracy of the prediction does not depend strongly
on the choice of $\tau$ (for a fixed total computational 
resource), again over a wide range of values.
However, when $\tau$ is too short, then fast growth
becomes susceptible to the problem of poor convergence
of exponential averages, familiar to any practitioner
of the free energy perturbation method (which in the
present context is the Widom particle insertion method).
Finally, we have considered the relationship between 
fast growth and the linear response approximation.

The natural question left open by our study is:
How generally valid are the results which we have
found empirically?
In particular, how robust is the conclusion about relative
performance of slow and fast growth, i.e.\ ``comparable 
accuracy for equal computational resource''?
If this conclusion is indeed generally true,
then it suggests certain advantages for fast growth
(easy estimation of errors, parallelizability).

Finally, it is likely that fast growth can be improved 
by combining it with strategies which have been 
studied in the context of the free energy
perturbation (Eq.\ref{eq:fep}).
Suggestions to this end have included 
the overlapping distributions method\cite{frenkel_pc}, 
higher-order cumulant expansions\cite{hummer_pc}, 
and averaging of forward and reverse processes\cite{hermans_pc}.

\section*{Acknowledgments}

The authors are grateful for stimulating discussions
with Tanmoy Bhattacharya, Daan Frenkel, Angel Garcia, Jan Hermans,
Gerhard Hummer, Lawrence Pratt, Mike Warren, and Robert Wood,
as well as scrupulous and useful commentary by an anonymous
referee.
This research is supported by the Department of Energy,
under contract W-7405-ENG-36.
D.A.H. furthermore acknowledges the support of the
1999 Los Alamos Summer School program.

\begin{references}
\bibitem{initialmicrostate}
Alternatively, the initial microstate could have been used.
\end{references}
\end{document}